\documentclass[10pt]{article}

\usepackage{amsmath}
\usepackage{amssymb}

\usepackage{graphicx}

\usepackage{cite}

\usepackage{color} 

\usepackage{soul}

\usepackage[labelfont=bf,labelsep=period,justification=raggedright]{caption}

\makeatletter
\renewcommand{\@biblabel}[1]{\quad#1.}
\makeatother

\date{}

\begin{document}

\begin{flushleft}
{\Large
\textbf{THE BUILD-UP OF DIVERSITY IN COMPLEX ECOSYSTEMS}
}
\\
Andrea Tacchella$^{1}$,
Riccardo Di Clemente$^{2}$, 
Andrea Gabrielli$^{1}$,
 L. Pietronero$^{1,3,4}$
\\
\bf{1} Istituto di Sistemi complessi ISC-CNR UOS Sapienza, Piazzale A. Moro 5 Roma Italy\\
\bf{2} Massachusetts Institute of Technology, Department of Civil and Environmental Engineering, Cambridge, 02139, MA, USA\\
\bf{3} Sapienza, Universit\`a di Roma, Piazzale A. Moro 5 Roma Italy\\
\bf{4} London Institute for Mathematical Sciences, 35a South St, Mayfair London United Kingdom\\
\end{flushleft}


\section*{Abstract}
\addcontentsline{toc}{section}{Abstract}
Diversity is a fundamental feature of ecosystems, even when the concept of ecosystem is extended to sociology or economics. Diversity can be intended as the count of different items, animals, or, more generally, interactions. 

There are two classes of stylized facts that emerge when diversity is taken into account. Diversity explosions are the first stylized fact: evolutionary radiations in biology, or the process of escaping "Poverty Traps" in economics are two well known examples. 
The second stylized fact is nestedness: entities with a very diverse set of interactions are the only ones that interact with more specialized ones. In a single sentence: specialists interact with generalists. Nestedness is observed in a variety of bipartite networks of interactions: Biogeographic (Islands-Animals), macroeconomic (countries-products) and mutualistic (e.g. Pollinators-Plants) to name a few. This indicates that entities diversify following a pattern.

For the fact that they appear in such very different systems, these two stylized facts seem to point out that the build up of diversity might be driven by a fundamental mechanism of probabilistic nature, and in here we try to sketch its minimal features. Namely we show how the contraction of a random tripartite network, which is maximally entropic in all its degree distributions but one, can reproduce stylized facts of real data with great accuracy which is qualitatively lost when that degree distribution is changed.


We base our reasoning on the combinatoric picture that the nodes on one layer of these bipartite networks (e.g. animals, or products) can be described as combinations of a number of fundamental building blocks. We propose the idea that the stylized facts of diversity that we observe in real systems can be explained with an extreme heterogeneity (a scale-free distribution) in the number of meaningful combinations (\emph{usefulness}) in which each building block is involved. 
We show that if the \emph{usefulness} of the building blocks has a scale-free distribution, then maximally entropic baskets of building blocks will give rise to very rich behaviors in accordance with what is observed in real systems.


\section{Introduction}

The study of complexity in ecosystemic interactions has a long history. The seminal paper of May\cite{may} disputed the intuitive view that a complex network of interactions would tend to stability when its size is increased\cite{MACART}. May's formal results regarded networks with random interactions, but in real cases interactions are far from being random. 
In particular in this work we focus on the case of bipartite networks of interactions. In this case we can separate nodes in two layers such that nodes from one layer only have direct interactions with nodes from the other. This representation is useful when one layer can be interpreted as a set of possible resources for the nodes on the other one (with this relation being reciprocal in the case of mutualistic networks).


By enlarging the scope of the analysis, we can notice that bipartite networks of interactions are common in many fields such as biology (plants-pollinators\cite{bas1}, islands-species\cite{geobio1, atmar, darlington}), economics (countries-products\cite{natscirep,PlosOne2,HH2}, advertisement (customers-purchased items\cite{zhou2007bipartite,zhou2010solving})), sociology (sexual partners\cite{sexpartner}). The idea of considering economic or social interactions as entities affine to proper biological ecosystems isn't new. At the root of this association stands the fact that in both these two contexts there are entities competing and interacting for resource allocation. In such contexts being able to rely on a diversified set of resources is of course a great advantage, because it improves resiliency. At the same time exclusivity, namely having access to resources which are challenged by a small number of rivals, can boost this advantage. In a dynamic ecosystem fitter entities explore a phase-space of features (phenotypes in biology, or capabilities in economics) that allow them to make use of a possibly increasing range of exclusive resources. The result of such a dynamics can be a situation in which specialists (exclusive) resources are only accessible to fitter, generalists agents: a concept known as nestedness.

The idea that the same dynamics is taking place in such different contexts is strengthened by the observation that some stylized facts are present in observative data related to both economics and biologic ecosystems. 

The first stylized fact is related to a dynamic phenomena observed in complex ecosystems, namely sudden diversity explosions that are in sharp contrast with the previous rate of innovation of the system. In ecology this phenomena is known as evolutionary radiation. Examples of radiations are the well known Cambrian Explosion\cite{ECOCAMB}, or the evolution of insect-eating placental mammals into a wide variety of herbivores, flying mammals and marine mammals just after the Cretaceous\cite{FORTEY}. The mechanisms that drive such very fast increases in biodiversity are still object of discussion. While many ecologists focus on environmental causes\cite{OXY, CALCIUM}, some others indicate that possible causes should be related to the appearance, by chance, of some novel functional traits (like eyes\cite{EYES}) that trigger a chain of evolutions by opening new possibilities. More in general some point out that a \emph{complexity threshold} might have been crossed\cite{KAUF}. 

\begin{figure}
\centering
\includegraphics[scale=0.40]{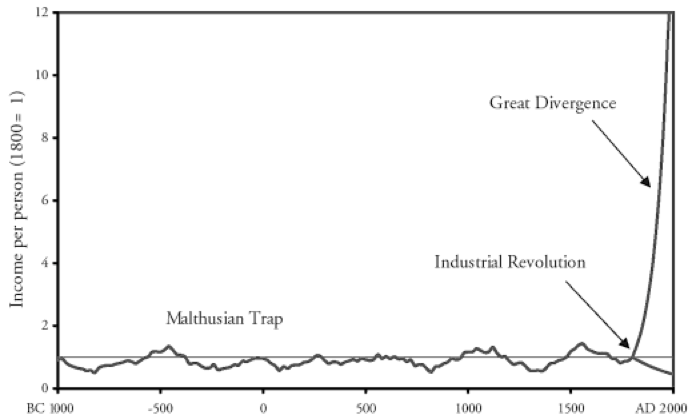}
\caption{The income per capita of the world population for the past 3000 years. The value remained more or less stable until the industrial revolution. Economy was so simple that whenever an improvement in productivity was made, the population simply grew accordingly, keeping the ratio between income and population constant. The industrial revolution marked a shift of regime. Interestingly it coincided with a divergence in which some countries remained stuck in the trap, while others escaped.}
\label{malthtrap}
\end{figure}

In economics something closely related has happened in relatively recent times.As a matter of fact, for most of human history the ratio between population and wealth remained constant (see fig. \ref{malthtrap}). This phenomena is known as Malthusian Trap\cite{malthtrap}. Malthus' theory was that population was merely limited by available resources and that at whatever point in time a technological progress was made, allowing access to greater resources, population would rapidly grow accordingly, thus keeping the wealth per capita in a trapped state. In the last years of Malthus' life the industrial revolution was beginning in England. The process lead to what is today known as The Great Divergence\cite{POMERANZ}: the industrialized countries were able to escape the Malthusian Trap and the wealth per capita has since then grown tremendously in these countries. In other words the industrial revolution coincided with a real shift of regime. 

If we consider the industrial revolution from the point of view of technological diversity it is not very different from an evolutionary radiation. In an incredibly short amount of time a variety of new technologies, resources, scientific advances and consumable products has stemmed, as the result of the introduction of a single new idea, the motor, in an ecosystem of technologies that weren't combined with that efficiency before. It was likely not the first time in history that such kind of revolutions happened (one can think of the invention of the wheel, or the development of agriculture) and the interesting fact is that it was still happening in recent times. In fact, in the recent past, just after the second World War, many countries were still living in a Malthusian Trap (or Poverty Trap). Some of these countries (e.g. China, India or South Korea) were able to escape the trap and move on to an industrialized rather than subsistence economy. In the process the diversity and complexity of their production exploded and this anticipated\cite{cloud} the later observed GDP growth.

\begin{figure}
\centering
\includegraphics[scale=0.40]{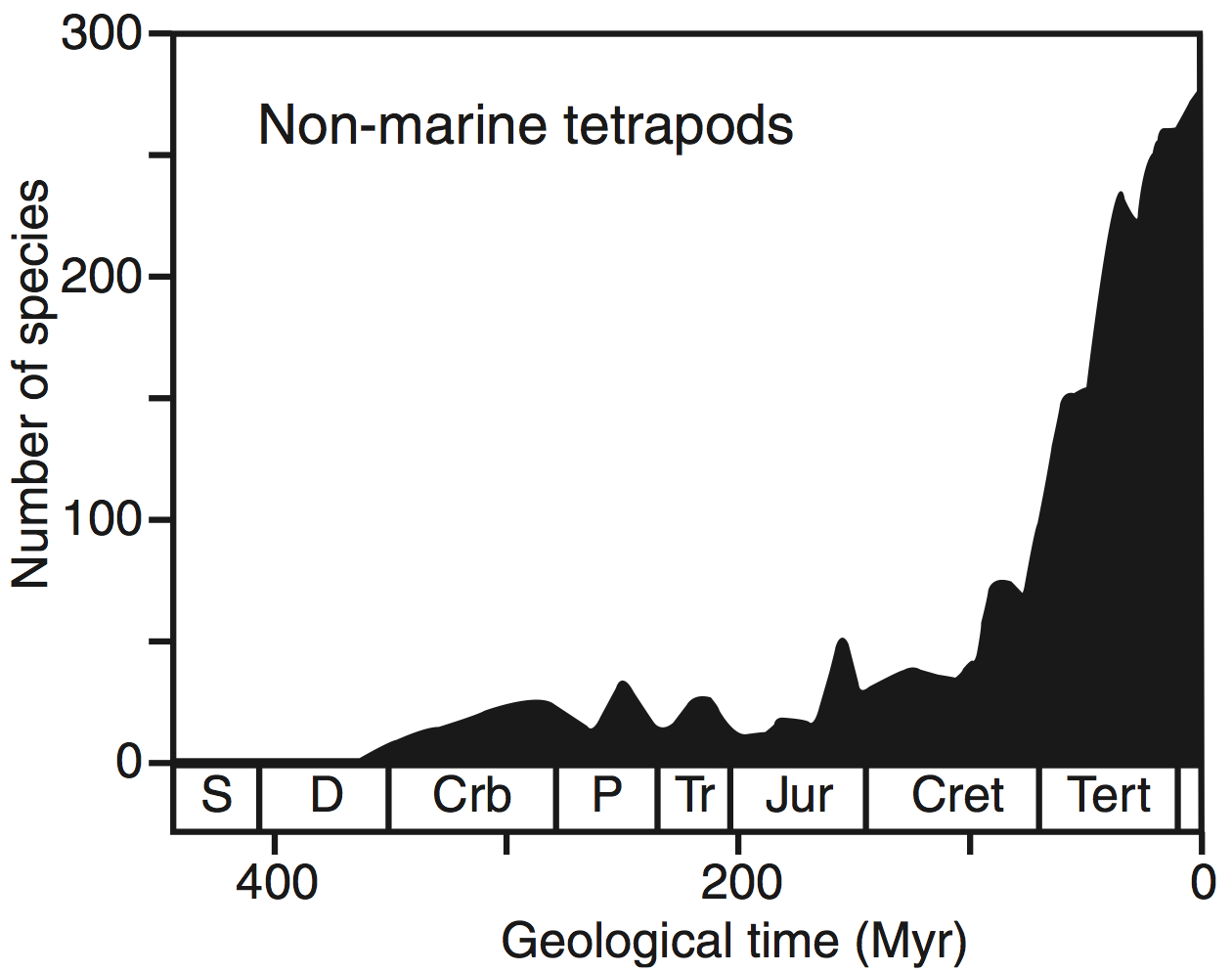}
\caption{The evolution of biodiversity of non-marine tetrapods. The dynamics is characterized by some interesting qualitative features. When we look at the general trend the growth is logistic, with an exponential increase which seems to be saturating. But at a finer level the dynamics shows a peculiar "bumpy" behavior, with bursts of diversity followed by periods of null growth or even extinctions.}
\label{tetrapods}
\end{figure}

It must be stressed that these topics are treated in a very qualitative way in literature for what concern the dynamics of diversity, both in economics and ecology. This is of course related to the fact that such revolutions happen rarely. Thus their characterization is hard to formalize and digging into the details of such dynamics is more a philosophical exercise rather than scientific. But yet the observation of these dynamics poses interesting scientific questions when one is interested in the general features. A good example is shown in fig. \ref{tetrapods}. The dynamics is characterized by some interesting qualitative features. When we look at the general trend the growth is logistic, with an exponential increase which seems to be saturating. But at a finer level the dynamics shows a peculiar "bumpy" behavior, with bursts of diversity followed by periods of null growth or even extinctions.
Given the rather qualitative level of such analysis, the introduction of a second class of stylized facts could help us settle the problem and identify some fundamental features of the build up of diversity.

The second stylized fact is known as nestedness in ecology and has been studied for a long time. In bipartite networks of interactions, nestedness is a peculiar degree correlation between nodes in the two layers: entities with a very diverse set of interactions are the only ones that interact with more specialized ones. In other words, specialists interact with generalists. It has been suggested that this organized structure of interaction may be beneficial for the stability of the ecosystem\cite{bas1, MARITAN}. Interestingly nestedness is also a very clear feature of international trade bipartite networks: in this case non-ubiquitous products are produced only by diversified countries, and non diversified countries only produce ubiquitous products. This feature of the countries-products network motivated the introduction and definition of the metrics described in \cite{PlosOne2}


Remarkably the same metrics can be applied to biological networks and yield a measure of the importance of a given node in relation to cascades of extinctions much more accurate than any other standard measure of centrality\cite{MUNOZ}. Interestingly the rankings given by the metrics seem to solve the long standing problem of the optimal nested ordering of a matrix in an almost optimal way, much more efficiently than the standard "nestedness temperature" approaches\cite{nesttemp}, which were already known to be problematic\cite{nesttempprobl}.
Nestedness emerges when we consider the diversity associated with multiple entities that evolved in the same ecosystem. The fact that diversity is organized suggests that the process of emergence of diversity follows a pattern. This consideration combined with the observation of similar stylized facts in systems of very different nature seem to point out that the build up of diversity might be driven by a fundamental mechanism of probabilistic nature.



\section{A model for the dynamics of diversity}

We picture diversity as the number of meaningful combinations of small pieces, or building blocks, that combine together to create meaningful associations. An ecosystem is thus a basket of such small pieces and the environment (and the competition) define which combinations are meaningful. The active agents of the ecosystem collect some of these building blocks from the basket and their fitness is larger the larger the number of meaningful combinations they can make, being larger the set of possible interactions they can have.
Thus, in the economic framework, a country will be able to produce all the products for which it owns all the needed building blocks, or capabilities. The concept of capability was introduced by Lall\cite{lall} and is at the basis of the idea of Economic Complexity. In this view capabilities are all the technical, political, geographical, infrastructural and social requirements that allow the production of a given product in a country. 
From a biological point of view we can consider the islands of an archipelago. An island will contain all the life forms that can emerge out of the genetic traits that are present in its ecosystem. In an archipelago, the fact that these genetic traits all come from a common pool (the basket) generates nestedness. In the case of mutualistic networks animals will be able to gather resources from all the plants that the combinations of their phenotypic traits will allow, and interestingly, the same can be said for plants in the opposite direction.


In this view, the observed diversity is the result of two processes. The first is the random appearance of novel traits (or technologies, or ideas), with diversity increasing as the number of new combinations made possible. The second is natural selection: non-meaningful o non-fit combinations are removed from the system, thus decreasing diversity. In this work we focus on the first process and we schematize the second in a statical way. In particular we assume that the natural selection is always at equilibrium in our model, and we simply impose a set of acceptable combinations and discard all the others, with the requirement that the number of acceptable combinations is much smaller than the number of possible combinations.


\subsection{Explosions of diversity and the concept of \emph{Usefulness}}
\begin{figure}
\centering
\includegraphics[scale=0.48]{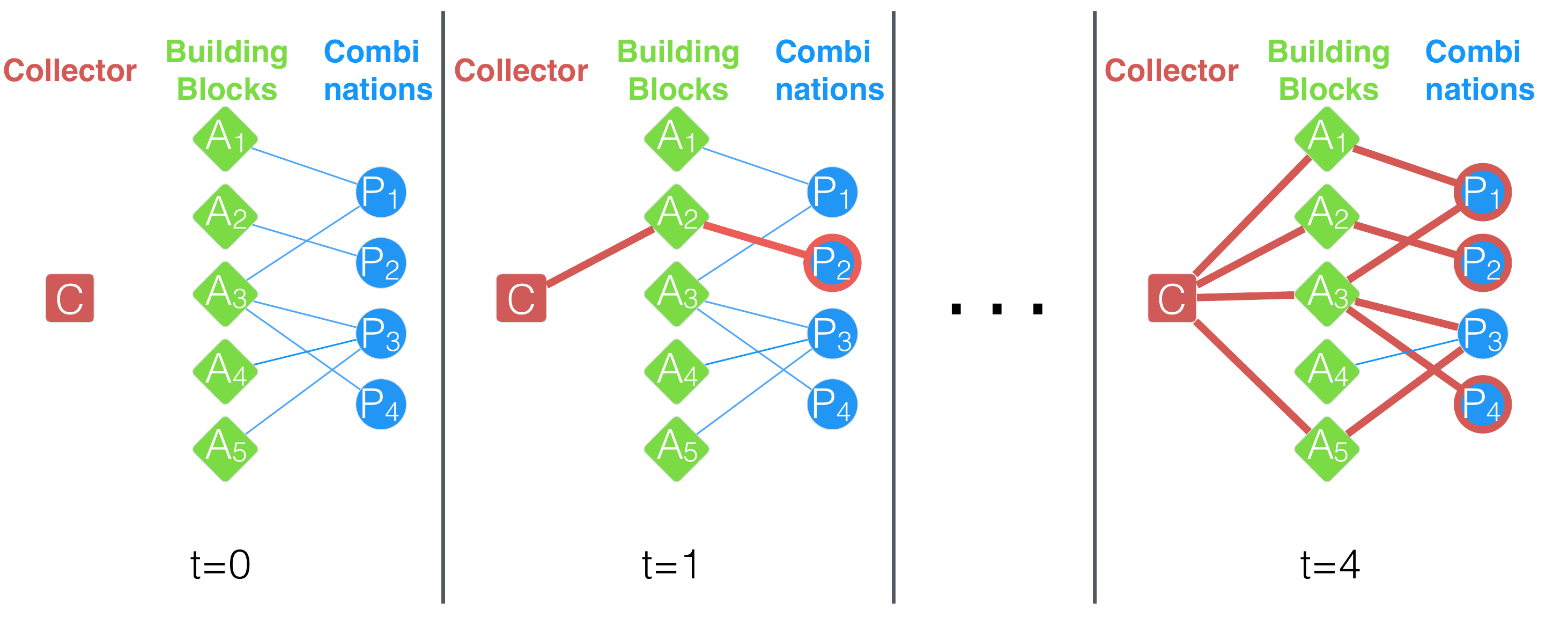}
\caption{A schematics representation of the framework of the model: a bipartite network connecting Building Blocks and combinations is defined and a collector gathers Building Blocks through time. The dynamics is implemented as follows: at each time step we add a link between the Collector and one randomly chosen Building Block. The diversity at that time step is the number of Combinations that the Collector is able to make given the Building Blocks it has collected until that time step.}
\label{dynmodel}
\end{figure}

First we try to characterize a minimal model that is able to reproduce qualitatively dynamics close those shown in figg. \ref{malthtrap} and \ref{tetrapods}.
The framework of the model is easily understood by looking at fig. \ref{dynmodel}. Building blocks and combinations form a bipartite network. The collector is endowed with an increasing number of building blocks, one at a time. This correspond to the disputable hypothesis of a constant rate of exploration, which is in any case the simplest assumption in this framework. The diversity of the collector at any given time is the number of combinations for which it has all the building blocks.

What is left to define are the properties of the topology of the bipartite network of building blocks and combinations. Again, to keep the model minimal, we define this network to be random and we only focus on its degree distributions. In these degree distributions stands the key feature that embodies our view of the fundamental mechanism driving the build-up of diversity: the concept of \emph{Usefulness}. We define \emph{Usefulness} as the number of meaningful combinations in which a building block is involved. Previous implementations of similar models such as those presented in \cite{HH2} did not make use of this concept and implicitly imposed a binomial distribution for the \emph{Usefulness} of the building blocks. We argue that in nature this isn't the case. We can think of a number of practical situations in which this distribution would be substantially different than a unimodal exponentially decaying one: in technology ideas such as the transistor have clearly a much larger number of applications than, say, a particular metal working technique; in biology, as already noted, phenotypic features such as eyes find place on a much larger set of fit life forms than, say, the pouch of marsupials. One could also notice that the \emph{Usefulness} of wings would be somewhere in the middle among eyes and the pouch, but yet with orders of magnitude of distance from each of the two. A mathematical formalization of these ideas corresponds to a "fat-tailed" distribution of \emph{Usefulness}, and as we show this assumption allows for a clear qualitative shift in the output of our models.

In detail we build the bipartite Building Blocks-Combinations network as follows:
\begin{itemize}
\item{First we draw from a power-law distribution $P(n)\propto n^{-\alpha}$ a number $n_i$ for each building block, that is its \emph{Usefulness}.}
\item{Then for each Building Block $i$ we choose randomly $n_i$ Combinations to which the Building Block will be connected. In this way the expected value of the length of each Combination is the same.}
\end{itemize}

As mentioned the dynamics is implemented in the simplest possible way: at each time step we add a link between the Collector and one randomly chosen Building Block. The diversity at that time step is the number of Combinations that the Collector is able to make given the Building Blocks it has collected until that time step. About the numerosity of the nodes of the bipartite network we only request that the number of possible combinations of Building Blocks is much larger than that of the actually allowed ones. In practice all the results shown here 
are obtained with a fixed number of Building Blocks $N_a$ and Combinations $N_p$ with $N_a=N_p=1131$. Changing these numbers and the ratio between $N_p$ and $N_a$ only causes quantitative changes in the behaviors, as long as the two numbers are large enough and as long as the number of possible combinations of Building Blocks remains much larger than that of the allowed Combinations.

\begin{figure}
\centering
\includegraphics[scale=0.48]{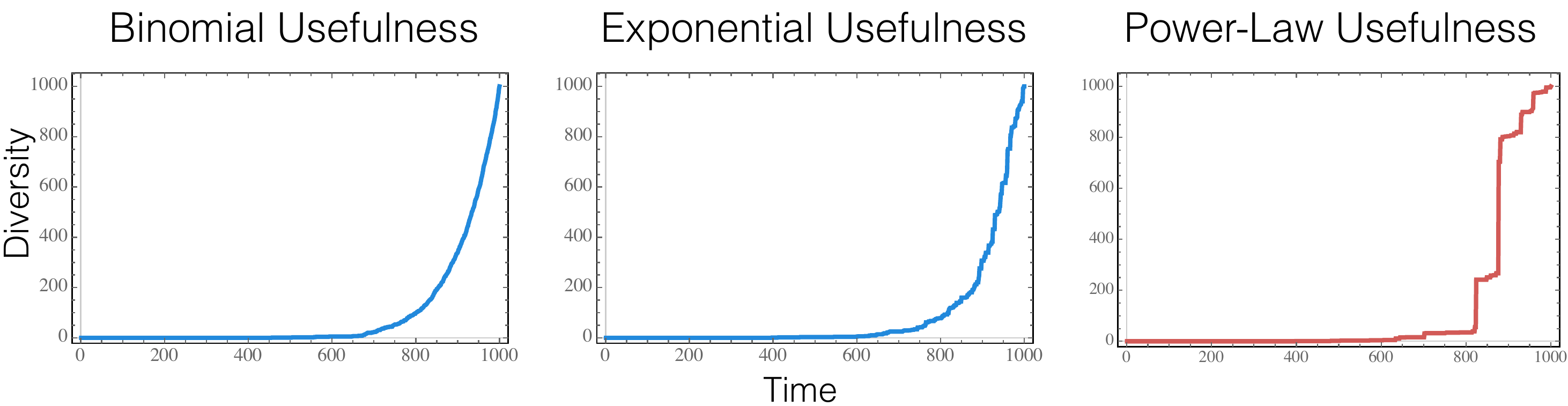}
\caption{A schematics representation of the framework of the model: a bipartite network connecting Building Blocks and combinations is defined and a collector gathers Building Blocks through time. The dynamics is implemented as follows: at each time step we add a link between the Collector and one randomly chosen Building Block. The diversity at that time step is the number of Combinations that the Collector is able to make given the Building Blocks it has collected until that time step.}
\label{dynamics}
\end{figure}

In fig. \ref{dynamics} we show the resulting dynamics for three possible distributions of \emph{Usefulness}. The first two behaviors show an essentially exponential increase in diversity. While the loss of the unimodality for the exponentially decaying distribution causes the trajectory to be a bit rougher, it is clear the qualitative shift that is obtained with the slowly decaying distribution. The global "logistic-like" shape of the curve and the bursts of activity shown in the biological example of fig. \ref{tetrapods} are qualitatively well reproduced. What is missing is of course the extinction part, which is out of the scope of this model, since, as we stated, we consider the natural selection to be always at equilibrium, thus unfit Combinations are automatically suppressed.

We can interpret the dynamics with the presence of two regimes: at the beginning the system lives in the "poverty trap", when for a long time no relevant increase in diversity is observed. Then, once the system has accumulated a large enough number of Building Blocks, or complexity, the dynamics shift to a different regime of fast growth. The sudden increases in diversity corresponds to the discovery of a very \emph{Useful} idea, that allows to exploit a large part of the Building Blocks already owned but that were missing a piece that could tie them together.


\section{Many collectors at the same time: emergence of complex nestedness}
The bipartite networks in which we observe nestedness can be put in correspondence with a generalization of the dynamical model proposed in the previous section, but observed at a fixed time step. In particular it correspond to consider many independent collectors that are endowed with random sets of Building Blocks, drawn from the same basket. Thus all the collectors are in principle the same, and no heterogeneity, other than different random choices of the capabilities is introduced. As we will see this will nevertheless result in a significant heterogeneity of the resulting diversity among the Collectors, when we consider the contracted bipartite networks of Collectors with the Combinations that they can make

\begin{figure}
\centering
\includegraphics[scale=0.48]{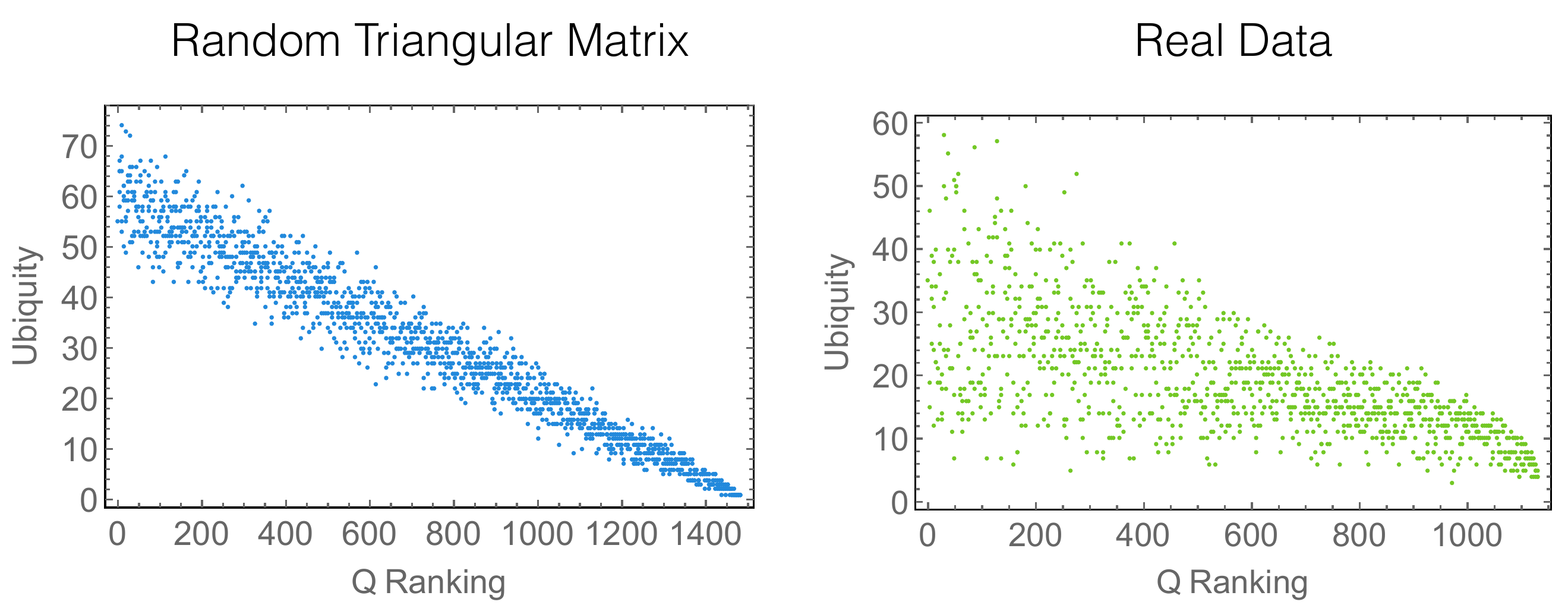}
\caption{Ubiquity versus Complexity ranking for a random triangular matrix and the real $M_{cp}$ for year 2010. In the random case the knowledge of ubiquity is almost the same as the ranking of complexity. In the real case the situation is much different. Ubiquity is not a good proxy for complexity
since even a non ubiquitous product can be of low complexity if a non-diversified country is able to export it.  This complexity is not present in the random case.}
\label{randtriang}
\end{figure}

For the fact that we developed quantitative methods to describe the properties of the adjacency matrix of the bipartite Collectors-Combinations projection of such network, we can describe in more detail some stylized fact present in real data. From a qualitative point of view we can begin with the trivial observation of the shape of the matrix: once rows and columns are ordered by Fitness and Complexity the shape is triangular-like. While this is itself a symptom of nestedness it is worth to notice that there is something more about the structure of these matrices that we can observe. We can, for example, build a random binary matrix in which a given density of 1's is concentrated in one of its two triangles while the other is left empty. The shape is now qualitatively similar to that of the real matrix. Nevertheless, we can use Fitness and Complexity to spot that this randomness does not reproduce the full complexity of the real data. In fig. \ref{randtriang} is shown a comparison of the Ubiquity versus Complexity Ranking plot for a random triangular matrix and the real $M_{cp}$ for year 2010. In the random case the knowledge of ubiquity is almost the same as the ranking of complexity. In the real case the situation is much different. Ubiquity is not a good proxy for complexity
since even a non ubiquitous product can be of low complexity if a non-diversified country is able to export it. This complexity is not present in the random case. Thus we start to understand that the matrix is not only nested, but some non trivial structure, or complexity, is present in its fine details.

\begin{figure}
\centering
\includegraphics[scale=0.48]{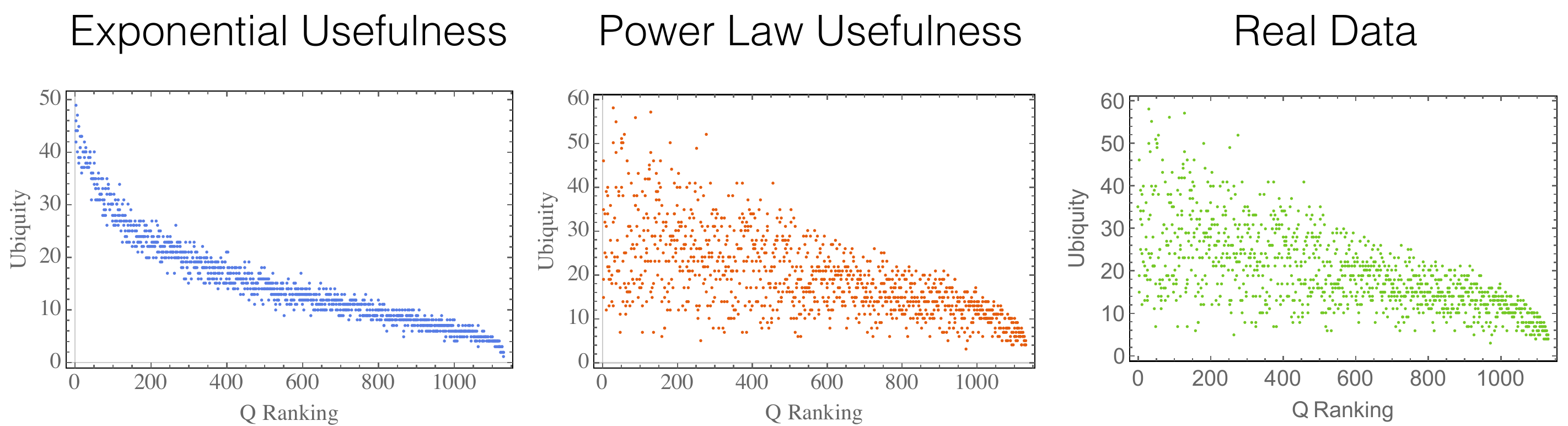}
\caption{Ubiquity versus Complexity ranking for the cases of different distributions of \emph{Usefulness} and the real data. As it is clear, the power-law distribution gives results in good accordance with the real case.}
\label{qvsdiv}
\end{figure}

We can then try to use Fitness and Complexity to assess the accuracy with which our models are able to reproduce not just the shape, but also the finer features of the nested matrices that we observe. 
In fig. \ref{qvsdiv} we show a comparison of the same plot for two different distributions of \emph{usefulness} and the real matrix.

As it is clear from the figure, the introduction of a "fat-tailed" distribution for the \emph{Usefulness} introduces a qualitative change in the fine structure of the $M_{cp}$ matrix.

\begin{figure}
\centering
\includegraphics[scale=0.48]{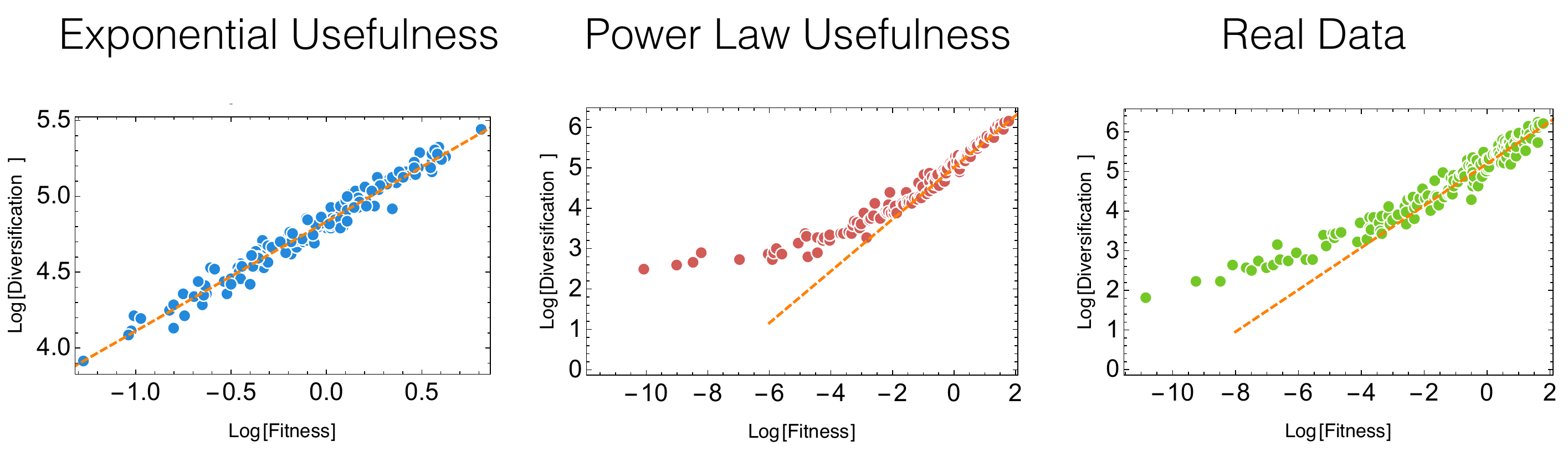}
\caption{Fitness versus diversification for two different distributions of \emph{Usefulness} and the real data. From the real data two regimes emerge in a clear way: collectors with low fitness live in a "Poverty Trap" where a given increase of complexity leads to a small increase in diversity; collectors with higher fitness (along the dashed trend line) have a much larger benefit from the same increase in complexity. Their efforts are thus much more rewarded. Interestingly the same two regimes are present in the power-law case but not in the exponential case. This is also reflected by the features of the dynamics shown in fig. \ref{dynamics}.}
\label{ptrap}
\end{figure}

Moreover we can use the Fitness to see what is the relation between the complexity of collectors and their diversification when different distributions of \emph{Usefulness} are considered. In fig. \ref{ptrap} we show the Fitness versus diversification plot for two different distributions of \emph{Usefulness} and the real data. From the real data two regimes emerge in a clear way: collectors with low fitness live in a "Poverty Trap" where a given increase of complexity leads to a small increase in diversity; collectors with higher fitness (along the dashed trend line) have a much larger benefit from the same increase in complexity. Their efforts are thus much more rewarded. Interestingly the same two regimes are present in the power-law case but not in the exponential case. This is also reflected by the features of the dynamics shown in fig. \ref{dynamics}.

\begin{figure}
\centering
\includegraphics[scale=0.48]{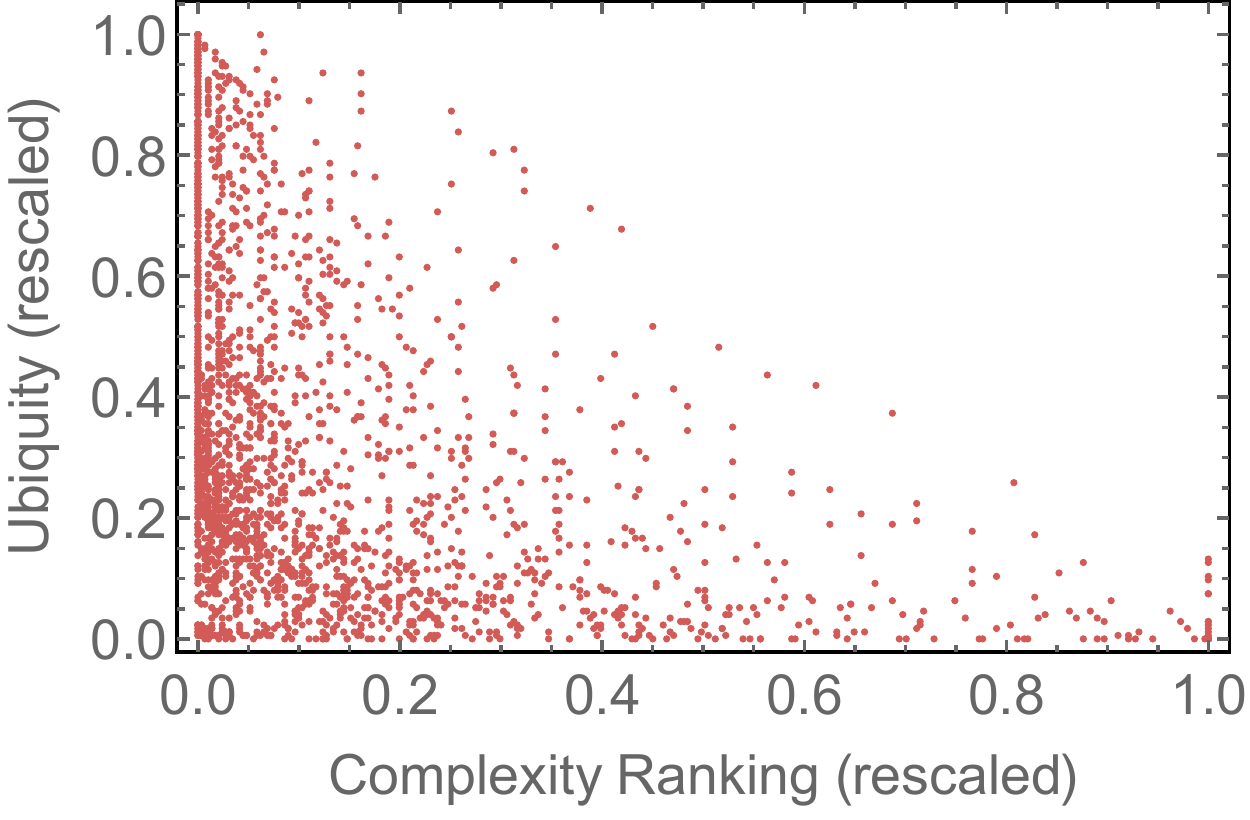}
\caption{Ubiquity vs. Complexity ranking relation for 59 Plant-Pollinators networks. The points are rescaled to collapse in a 1*1 box, being the size of the networks very heterogeneous.}
\label{plpol}
\end{figure}

We can also see how nested matrices from biological datasets display the same properties. As an example we plot the Ubiquity vs. Complexity ranking relation for 59 Plant-Pollinators networks\footnote{Source: Web of Life database (www.web-of-life.es)}, and the results are shown in fig. \ref{plpol}. Again the ubiquity is substantially different from complexity, in a way that only a fat-tailed distribution of \emph{Usefulness} is able to explain in this framework.

Even when comparing more quantitative features for real nested matrices the power-law distribution of \emph{Usefulness} proves to be able to give results in striking accordance with the real observations. Since all the other features of the tripartite Collectors-BuildingBlocks-Combinations network are random, we think that from this simple model we can learn something interesting about the mechanisms governing the build-up of diversity in complex ecosystems. 

\section{Fat Tailed distributions of usefulness in real data}
\begin{figure}
\centering
\includegraphics[scale=0.48]{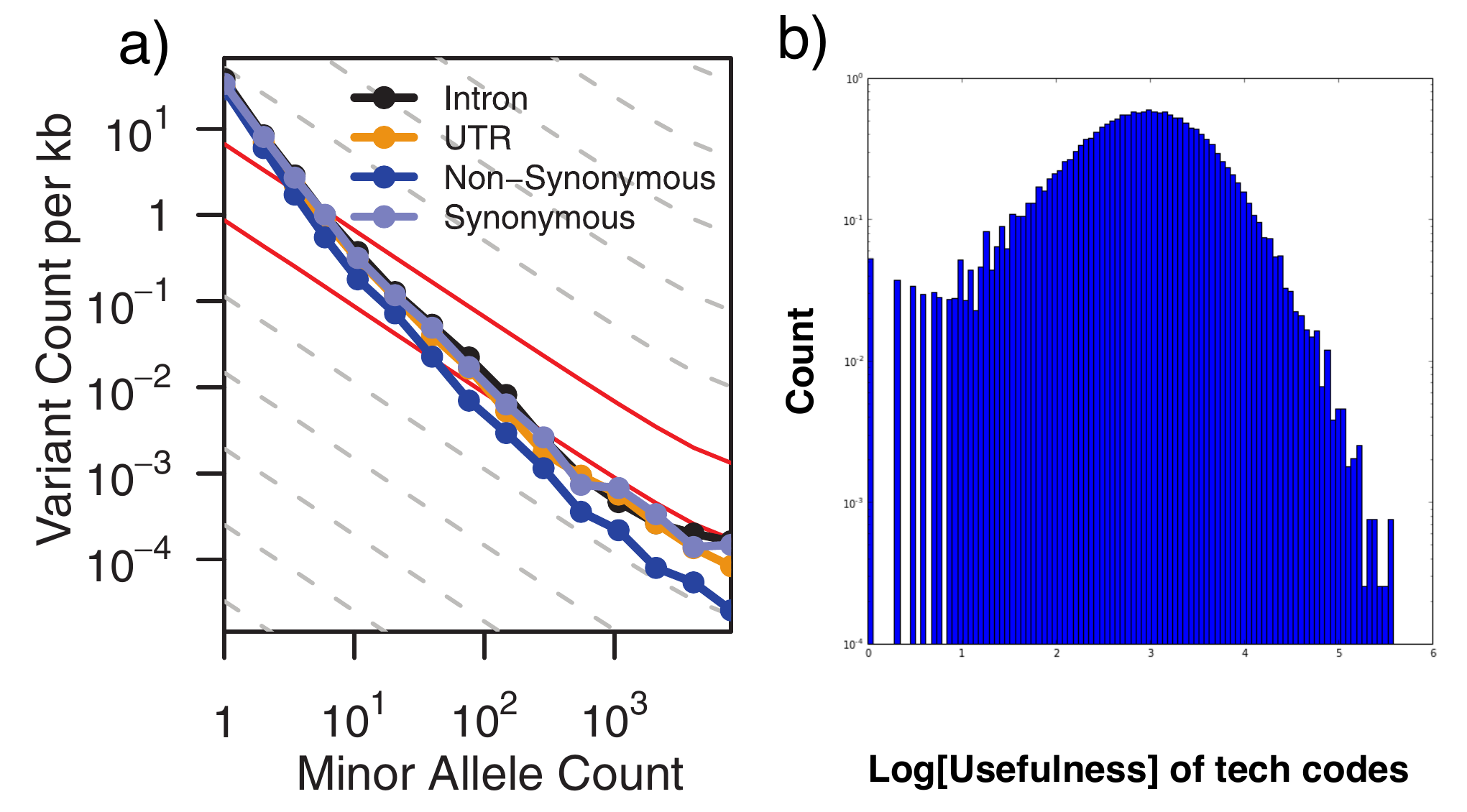}
\caption{a) Frequency of appearance of genetic variants in a population of 15000 humans(From \cite{RAREGENES}). b) Frequency of appearance of technological codes in a dataset of patents (From \cite{napolitano})}
\label{realusef}
\end{figure}

It would be interesting to observe something similar to a distribution of \emph{Usefulness} in a real system. It is not easy to check experimentally if such kind of distribution do exist in nature, mostly because giving a precise definition of the Building Blocks is a hard task. In the technological case the qualitative observation that some technological ideas have found a much wider application than many other is readily made. A quantitative approach can be made with respect to the frequency with which we observe technological codes in patents: this seems to follow a roughly log-normal distribution. This "fat tailed" distribution corresponds to a non negligible number of technology codes appearing on a very large amount of patents. From a biological perspective some hints that these distribution are present in nature can come from genetics. Recent studies have demonstrated that the frequency with which genetic variants appear in a population follow a scale-free distribution\cite{RAREGENES}. These findings are summarized in fig. \ref{realusef}.

\section{Conclusions}
By imposing very general conditions on a very simple model we try to highlight how the concept of \emph{Usefulness} might be crucial to understand the mechanism that drive the build-up of diversity in complex ecosystems. These fundamental mechanisms seem to be very general, as the stylized facts that they produce are observed in a wide class of systems, driven in principle by very different laws. The ideas that we propose in this 
work are anyway general enough to be transposed to different contexts with an obvious correspondence of the variables.

\bibliographystyle{plos2009}
\bibliography{Bibliography}

\end{document}